\documentclass[fleqn,usenatbib]{mnras}

\usepackage{newtxtext,newtxmath}
\usepackage{comment}
\usepackage{deluxetable}

\usepackage[T1]{fontenc}

\DeclareRobustCommand{\VAN}[3]{#2}
\let\VANthebibliography\thebibliography
\def\thebibliography{\DeclareRobustCommand{\VAN}[3]{##3}\VANthebibliography}

\usepackage{graphicx}	
\usepackage{amsmath}	
\usepackage{amssymb}	

\newcommand{\CII}{C~{\sc ii}}
\newcommand{\CIII}{C~{\sc iii}}
\newcommand{\CIV}{C~{\sc iv}}

\newcommand{\HI}{H~{\sc i}}

\newcommand{\lya}{Ly$\alpha$}
\newcommand{\lyb}{Ly$\beta$}
\newcommand{\FeII}{Fe~{\sc ii}}

\newcommand{\OI}{O~{\sc i}}

\newcommand{\OVI}{O~{\sc vi}}

\newcommand{\MgII}{Mg~{\sc ii}}
\newcommand{\MgI}{Mg~{\sc i}}

\newcommand{\NV}{N~{\sc v}}

\newcommand{\SiIV}{Si~{\sc iv}}

\newcommand{\kms}{\hbox{km~s$^{-1}$}}
\newcommand{\cmsq}{\hbox{cm$^{-2}$}}

\newcommand{\flux}{\hbox{erg~cm$^{-2}$~s$^{-1}$}}

\newcommand{\lumin}{\hbox{erg~s$^{-1}$}}

\newcommand{\nh}{\hbox{${N}_{\rm H}$}}

\newcommand{\be}{\begin{equation}}
\newcommand{\ee}{\end{equation}}
\newcommand{\ba}{\begin{eqnarray}}
\newcommand{\ea}{\end{eqnarray}}

\newcommand{\chandra}{{\emph{Chandra}}}

\newcommand{\hst}{\emph{HST}}

\newcommand{\simgt}{\lower 2pt \hbox{$\, \buildrel {\scriptstyle >}\over {\scriptstyle\sim}\,$}}
\newcommand{\simlt}{\lower 2pt \hbox{$\, \buildrel {\scriptstyle <}\over {\scriptstyle\sim}\,$}}
\newcommand{\ls}{\lower 2pt \hbox{$\;\scriptscriptstyle \buildrel<\over\sim\;$}}
\newcommand{\gs}{\lower 2pt \hbox{$\;\scriptscriptstyle \buildrel>\over\sim\;$}}
\newcommand{\sarc}{$^{\prime\prime}\!\!.$}

\newcommand{\bone} {B\,1152+199}

\title[Identifying the Lens Galaxy B\,1152+199 as a Ghostly Damped Lyman $\alpha$ System]{Identifying the Lens Galaxy B\,1152+199 as a Ghostly Damped Lyman Alpha System by the Cosmic Origins Spectrograph}

\author[Dai, Bhatiani, \& Chen (2020)]{
Xinyu Dai,$^{1}$\thanks{E-mail: xdai@ou.edu}
Saloni Bhatiani,$^{1}$
Bin Chen$^{2}$
\\
$^{1}$Homer L. Dodge Department of Physics and Astronomy, University of Oklahoma, Norman, OK 73019, USA\\
$^{2}$Research Computing Center, Department of Scientific Computing, Florida State University, Tallahassee, Florida 32306, USA\\
}

\date{Accepted XXX. Received YYY; in original form ZZZ}

\pubyear{2020}

\begin{document}
\label{firstpage}
\pagerange{\pageref{firstpage}--\pageref{lastpage}}
\maketitle

\begin{abstract}
Strong quasar-galaxy lensing provides a powerful tool to probe the inter-stellar medium (ISM) of the lens galaxy using radiation from the background quasar.
Using the Cosmic Origins Spectrograph (COS) on board the Hubble Space Telescope, we study the cold ISM properties of the lens galaxy in B\,1152+199 at a redshift of $z=0.4377$.
Since existing optical extinction and X-ray absorption measurements of the lens have revealed a large amount of cold ISM, we expected to detect a damped \lya\ absorption (DLA) system in the near ultraviolet spectrum; however, our upper limit on the \HI\ column density is several orders of magnitude below the expectation. 
We also marginally detect \OI\ and \CII\ absorption lines associated with the lens galaxy in the COS spectrum.
Thus, the lens galaxy is identified as a ghostly DLA system, and further investigations of these ghostly DLA systems would be important to characterize the biases of using DLAs to probe the matter density of the universe.
Although preliminary, the most likely explanation of the non-detection of the DLA is because of the \lya\ emission of the lens galaxy that fills in the absorption trough, with a \lya\ luminosity of $4\times10^{42}\,\lumin$.
\end{abstract}





\section{Introduction}

The inter-stellar medium (ISM) has a primary role in many areas of astronomy, including stellar formation/evolution, galaxy formation/evolution, physics of active galactic nuclei, and cosmology.
Quasar-galaxy strong gravitational lensing systems form a powerful tool to study the ISM properties of intermediate redshift galaxies.
Here, a bright background quasar lies with a small impact parameter behind a foreground galaxy (most lens galaxies are elliptical), and multiple images form due to the gravitational lensing effect. 
Because of the small impact parameter, the quasar light traverses the ISM of the lens galaxy, and the cold ISM leaves signatures on the lensed quasar
spectrum, including reddening due to dust, absorption of
the X-ray spectrum, and absorption lines in the ultraviolet (UV) and optical bands.  
This method is unique in its ability to probe ISM of high density regions deep within the lens galaxy.
Most absorption line systems probe lines of sight far more distant from the galaxy center \citep[e.g.,][]{ph04, menard09}.
Future large-scale surveys, such as the Large Synoptic Survey Telescope, will find thousands of strong quasar-galaxy lenses \citep{om10} that can be used for ISM studies.

The first two effects, dust extinction and X-ray absorption, have been investigated in the literature. 
For example, extinction laws of intermediate redshift galaxies 
has been studied in many gravitational lenses \citep[e.g.,][]{nadeau91,falco99,toft00,motta02,wucknitz03,munoz04,med05}.
Gravitational lenses also provide the first probes capable of studying the dust-to-gas ratios of cosmologically distant galaxies  \citep{dai03,dk05,dai06,dk09}.
By combining X-ray and optical measurements of both the gas and dust absorption between pairs of lensed images, we can estimate 
the dust-to-gas ratio $\Delta E(B-V)/\Delta\nh$ of the lens galaxies,
under the assumption that the differences between the extinction and gas absorption are due to the same parcel of ISM.  
\citet{chen13} found an evolving dust-to-gas ratio of $E(B-V)/N_H=1.17^{+0.41}_{-0.31} \times 10^{-22}\,\rm mag\,cm^2\,atom^{-1}$ in lens galaxies ($0<z<1$), with an intrinsic scatter of $\rm0.3\,dex$, lower than the Galactic value  $1.7\,\times 10^{-22}\,\rm mag\, cm^{2}\, atom^{-1}$ \citep{bsd78}, and a constant dust-to-metal ratio with redshift.
This paper focuses on the hydrogen and metal absorption lines in a UV spectrum of the lens galaxy in the two-image lens \bone.
\bone\ was selected as a lens candidate from the Cosmic Lens All-Sky Survey, and was confirmed as a two-image lens system with $z_s = 1.019$ and $z_l = 0.439$ by subsequent follow-up observations \citep{myers99}.  The \hst\ F555W and F814W band images of \bone\ show that the lensing galaxy may resemble an early-type galaxy \citep{rusin02}.
However, the large amount of cold ISM detected from the lens galaxy, through studying the absorption/extinction properties of the background quasar spectrum, will classify the galaxy as a late-type.
In particular, \citet{myers99} detected \MgII\ and \MgI\ metal absorption lines at the lens redshift from the optical spectrum of the quasar image A (C.\ Fassnacht, private communication). 
\citet{toft00} detected a large extinction for the system, especially in quasar image B, and 
 \citet{ela06} confirmed this result and measured a large
differential extinction of $\Delta E(B-V) = 1.20\pm0.05$ between images A and B with a slope of $R_V = 2.1\pm0.1$.
\citet{dk09} and \citet{chen13} detected cold ISM in the lens from X-ray spectra of the quasar images and measured differential absorption of $\Delta \nh = (4.8\pm0.4)\times10^{22}$~\cmsq\ assuming Solar metallicity.  Combining the differential extinction and absorption measurements, \citet{dk09} measured a dust-to-gas ratio of $E(B-V)/\nh = (2.5\pm0.2)\times10^{-22}$~mag~cm$^{2}$~atom$^{-1}$, slightly higher than the Galactic value $1.7\times10^{-22}$~mag~cm$^{2}$~atom$^{-1}$ \citep{bsd78}.
Since metals dominate the X-ray absorption cross section, the \nh\ measured from the X-ray spectrum is an effective column density rather than a direct measurement. Ideally a direct measurement of \nh\ using other methods can be vital to break the degeneracy in the X-ray analysis.
The original goal of our Cosmic Origins Spectrograph (COS) observation of \bone\ was to directly measure \nh\ through the damped \lya\ absorption (DLA) feature associated with the lens galaxy and also to detect other associated metal absorption lines in the brighter image A.   
Here, we present the non-detection of the DLA in \bone.  The observation and data products are introduced in Section\,2, followed by the imaging analysis in Section\,3 and the spectral analysis in Section\,4.  We discuss the results in Section\,5.
We assume cosmological parameters of $\Omega_M = 0.27$, $\Omega_{\Lambda}=0.73$, and $H_0 = 70$~km~s$^{-1}$~Mpc$^{-1}$ throughout the paper.

\begin{figure}
	\includegraphics[width=\columnwidth]{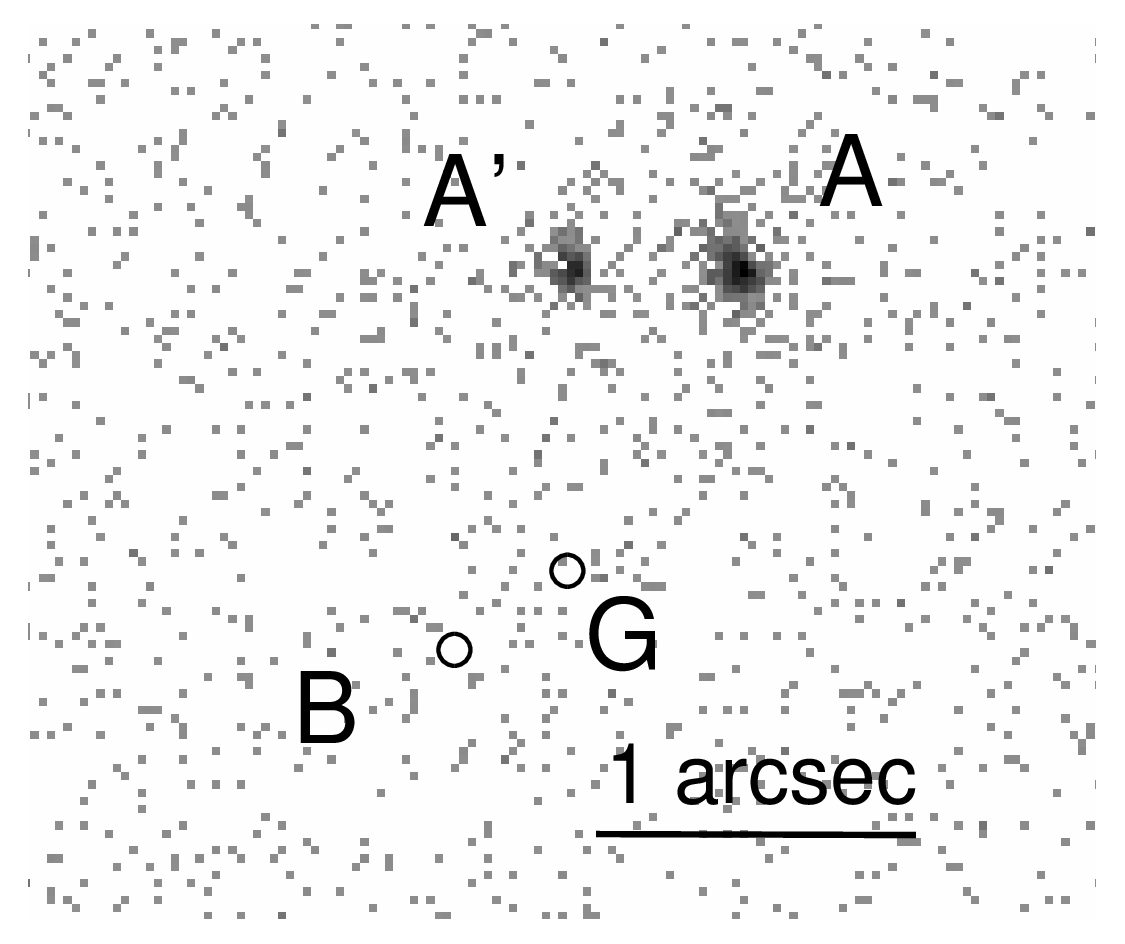}
\caption{\hst/COS acquisition image of \bone, where only image A is detected.  A' is the additional ghost image produced by MirrorB. The expected locations of image B and the lens galaxy are marked by circles. \label{fig:img} }
\end{figure}

\begin{figure}
	\includegraphics[width=\columnwidth]{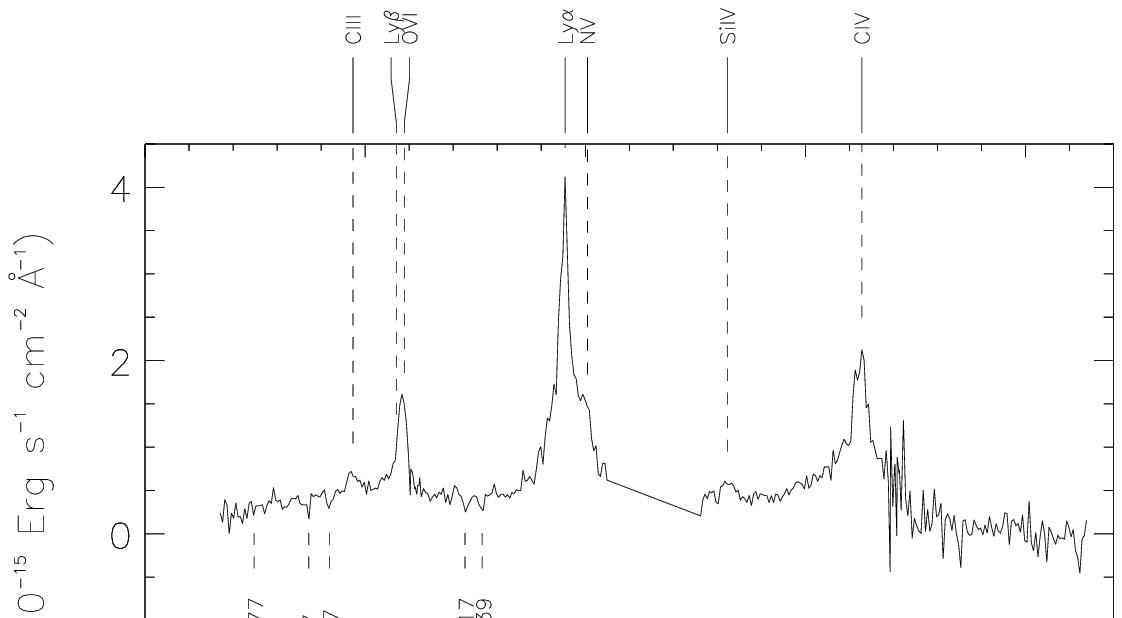}
\caption{\hst-COS-G230L NUV spectrum of \bone, binned by 5\AA.  \CIII, \lyb, \OVI, \lya, \NV, \SiIV, and \CIV\ emission lines from the background quasar at $z_s = 1.0155$ are labeled on top of the spectrum.  The expected strong \lya\ absorption feature from the lens galaxy $z_l = 0.4377$ is not detected.  Four low significance absorption lines are detected: two of them are \OI\ and \CII\ absorption lines associated with the lens galaxy, and the other two are other \lya\ absorbers along the line of sight at different redshifts. \label{fig:spec} }
    \label{fig:example_figure}
\end{figure}

\begin{figure}
	\includegraphics[width=\columnwidth]{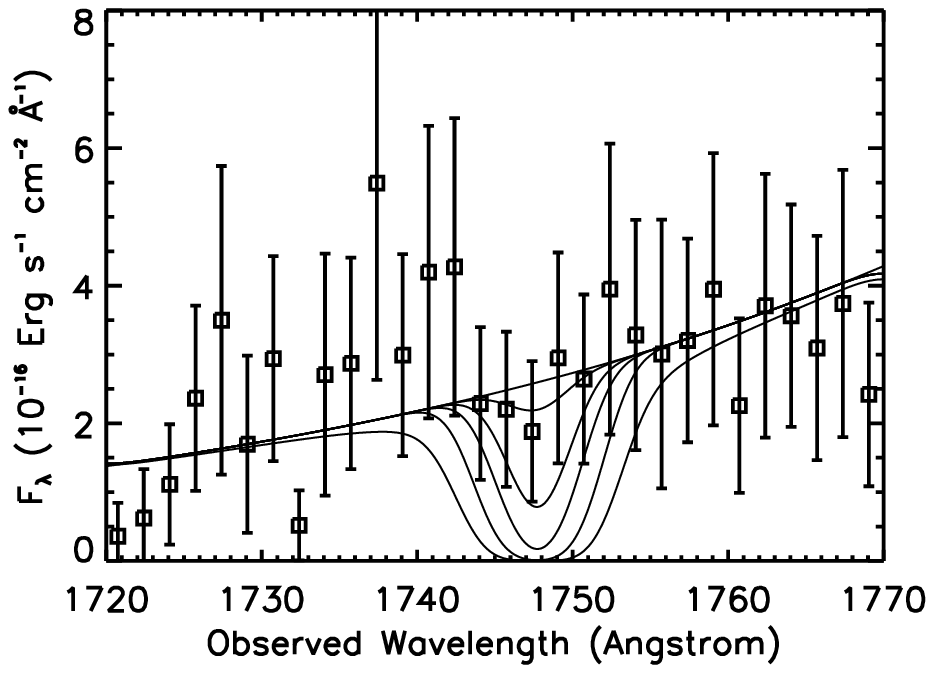}
\caption{The NUV spectrum (squares) between 1720--1770\AA\ of \bone.  A DLA system was expected at 1748.8\AA\ (observed frame), which is not detected in the spectrum.  The solid curves show a power law model absorbed by Voigt profiles with $\nh = 0$, $10^{14}$, $10^{15}$, $10^{16}$, $10^{18}$, $10^{20}\,\cmsq$, from the top to the bottom, further smoothed by the 1.5\AA\ resolution of the spectrum.  We measured a 1$\sigma$ upper limit for the neutral hydrogen column density of $\nh < 5.4\times10^{14}~\cmsq$. \label{fig:dla} }
\end{figure}

\section{Observations}
We obtained a low-resolution NUV spectrum of \bone\ using COS \citep{green12} onboard \hst\ with the G230L grating on 2014--04--15.
COS is a complicated instrument that can be configured to take different observations.  For the interest of this paper, the acquisition image has a dimension of 10$\times$20~arcsec$^2$ with a spatial resolution of 0\sarc05, and the G230L grating spectrum is slitless with a field of view of 1\sarc25 in radius and a resolving power between 1500--2900.
We used two central wavelengths to cover a broader wavelength range and four positions for each central wavelength to minimize any local systematics in the detector, such as the bad pixels.
The detailed observation log is listed in Table~\ref{tab:obs}.
The data products were retrieved from STScI after updating the wavelength calibration issue in the G230L grating mode in May 2016, and we used the standard pipeline-produced, background-subtracted spectrum in our subsequent analysis.

\section{Imaging Analysis}
We analyzed the acquisition image of \bone\ (Figure~\ref{fig:img}), taken with the COS/NUV/MirrorB configuration with an exposure of 95.6~sec, and only detected image A.  Although two point sources were detected, they are the double-image caused by MirrorB for one single point source.
The imaging analysis was performed with \verb+Sherpa+\footnote{The Sherpa website is at https://cxc.harvard.edu/sherpa.}, which is a part of the \chandra\ data analysis package CIAO for imaging and spectral analysis and is distributed separately as a stand-alone python package.
We first modeled the acquisition image within the central 2\sarc5 radius region using two Gaussian components (A and A') for the double-image of image A and a constant background, and obtained an acceptable fit with a cash statistic of 15972.9 over 31408 degrees of freedom (dof).
We then added two additional Gaussian components (B and B') for the double-image of image B to the model. The relative position between B and A is set by previous \hst\ measurements \citep{rusin02}, and the FWHMs, relative position, and flux ratio between B and B' are set to be the same as those for A and A'.  Essentially, we added one free parameter to the model, the normalization of B.  We, again, obtained an acceptable fit with a cash statistic of 15972.2 over 31407 dof, and
constrained the 3$\sigma$ upper limit of the flux ratio between image B and A as $f_B/f_A < 100$.
Based on the differential extinction measurement of $\Delta E(B-V) = 1.20\pm0.05$ between images A and B with $R_V = 2.1\pm0.1$ \citep{ela06}, we expect that image B is fainter than image A by 9--10~mag in the NUV band, consistent with our flux ratio limit.

\begin{figure}
	\includegraphics[width=1.0\columnwidth]{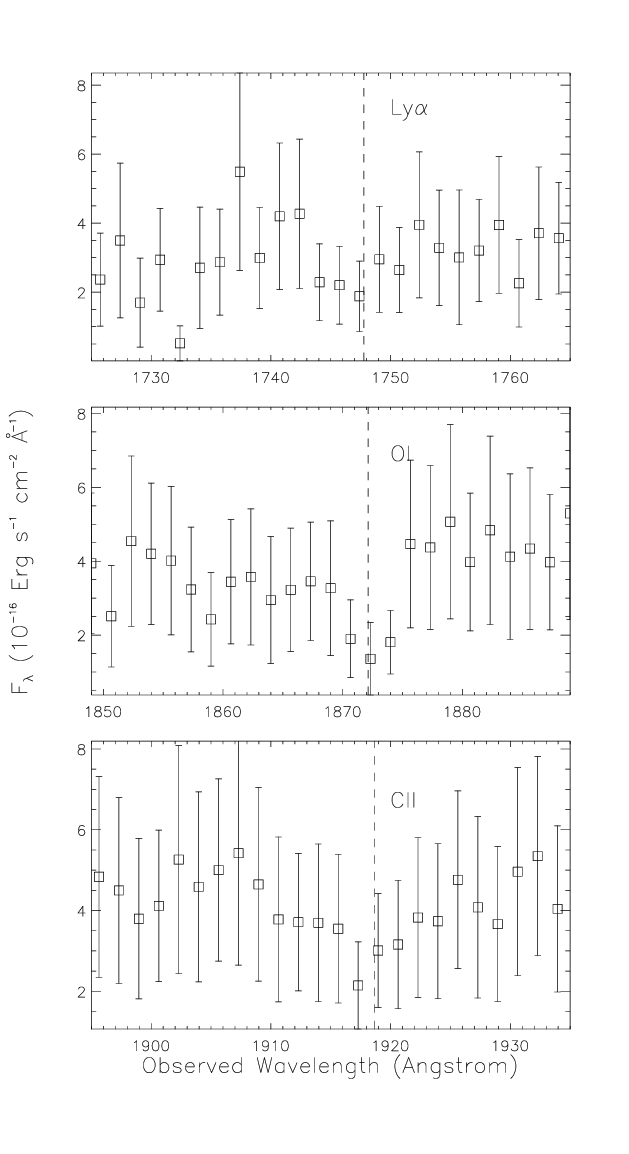}
\caption{Zoomed in NUV spectrum of \bone\ close to the absorbers associated with the lens galaxy.  The top panel shows the non-detection of the expected \lya\ absorption.  The middle and bottom panels show the \OI\ and \CII\ lines from the lens galaxy.  The vertical lines show the expected wavelengths of these lines for a lens redshift of $z_l=0.4377$. \label{fig:zoom}}
\end{figure}

\begin{figure}
	\includegraphics[width=\columnwidth]{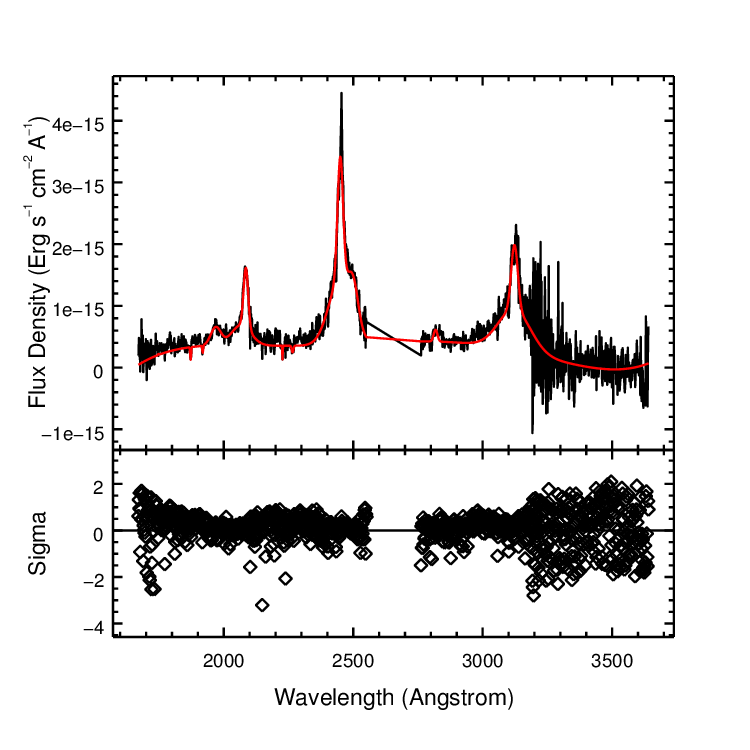}
\caption{\hst-COS-G230L NUV spectrum of \bone, binned by 1.5\AA, and its best-fit model (red line).  The model is composed of a six-component polynomial for the continuum emission, nine emission lines from the background quasar (\CIII, \lyb, \OVI, broad and narrower \lya, \NV, \SiIV, and broad and narrower \CIV\ lines), and five absorption systems. \label{fig:model} }
\end{figure}


\begin{table*}
\caption{\hst-COS Observations of \bone. \label{tab:obs}}
\begin{tabular}{cccccc}
\hline
\hline
Target & Instrument & Grating & Wavelength & Resolving & Exposure \\
& & & Coverage (\AA) & Power\tablenotemark{a} & Time (sec) \\
\hline
\bone & COS/NUV/MirrorB & None          & 1200--3300             & \nodata    & 95.6   \\
\bone A & COS/NUV         & G230L 2950\AA & 1650--2050, 2750--3150 & 1550--2900 & 7950.4 \\
\bone A & COS/NUV         & G230L 3360\AA & 2059--2458, 3160--3560 & 1550--2900 & 1853.3 
\\
\hline
\end{tabular}
\tablenotetext{a}{The spectral resolving power increases with the wavelength.}
\end{table*}


\section{Spectral Analysis}
Figure~\ref{fig:spec} shows the background subtracted NUV spectrum (binned by 5\AA) of \bone, a typical broad line quasar spectrum, to illustrate the main features of the spectrum.  Since image B is expected to be 9--10~mag fainter than image A in the NUV band, the spectrum is essentially that of image A.
The actual spectral fitting was performed on the observed spectrum binned by 1.5\AA\ (Figure~\ref{fig:model}), slightly larger than the resolution limit of 1\AA\ of the observation. 
We also used \verb+Sherpa+ to model the spectrum, and we first empirically fit the spectrum using a polynomial continuum model plus nine Gaussian emission lines.
We found that a six-component polynomial is sufficient to model the continuum, and the nine detected quasar emission lines are the broad \CIII, \lyb, \OVI, \lya, \NV, \SiIV, and \CIV\ lines along with narrower components for the \lya\ and \CIV\ lines.  
We set the width of the broad \lya\ and \lyb\ lines to be the same in all the fits.
We linked the wavelengths of these lines and jointly fit the redshift of the quasar as $z_s = 1.0155\pm0.0013$, with a fitting statistic of $\chi^2/dof = 578.4/1066$.
Table\,\ref{tab:em} lists the emission line properties.
Compared to the quasar redshift, $z_s = 1.0189\pm0.0004$, measured by Keck II LRIS in the optical band \citep{myers99}, the difference is 3.2$\sigma$ and we attributed the difference as the wavelength calibration difference between Keck II LRIS and \hst/COS.
The wavelength calibrations differ either by a multiplicative factor of 0.9966 or a linear shift of $3.56$\AA\ (observed frame).
We next searched for absorption lines in the spectrum.  The most significant feature (non-feature) of the spectrum in the absorption regime is the lack of any strong \lya\ absorption associated with the lens galaxy (Figure~\ref{fig:dla}).
Based on the extinction and X-ray absorption measurements of \bone, we expect a neutral hydrogen column density of $\nh \sim 10^{20}\,\cmsq$ for image A from the lens galaxy, which would result in a damped \lya\ line.
However, no absorption line is detected at the expected wavelengths 1745--1750\AA\ assuming that the lens redshift is between $z_l =$0.436--0.439.  
We estimated an upper limit on the neutral hydrogen column density by modeling the 1720--1770\AA\ spectral segment by a power-law modified by a Voigt absorption profile.  
We adopted an intrinsic velocity dispersion of 246\,\kms\ at the rest-frame of the absorber for the Gaussian component, based on a recent lens model constraint of the lens $\sigma=246^{+2}_{-9}$~\kms \citep{asadi20}, and the absorbed model was further smoothed by the 1.5\AA\ bin size of the observed spectrum.  The 68\%, 90\%, and 99\% confidence limits on the  neutral hydrogen column density are $\nh < 0.54$, 1.1, and 5.5$\times10^{15}$~\cmsq, respectively.
\begin{table}
\caption{Absorption Line Candidates in the COS UV Spectrum. \label{tab:em}}
\begin{tabular}{cccc}
\hline
\hline
Line & Redshift & Rest-Frame FWHM & Flux \\
& & \kms & $10^{-14}$~\flux \\
\hline
\CIII & 1.0155  & 3458$\pm$1080 & 1.4$\pm$0.4\\
\lyb &  \nodata\tablenotemark{a} & \nodata\tablenotemark{b} & 3.1$\pm$0.7\\
\OVI &  \nodata & 1490$\pm$380  & 2.15$\pm$0.63 \\
\lya &  \nodata & 6366$\pm$1900 & 11.8$\pm$2.8\\
\lya &  \nodata & 1517$\pm$846  & 5.2$\pm$2.8\\
\NV &   \nodata & 2685$\pm$1365 & 2.6$\pm$1.0\\
\SiIV & \nodata & 894$\pm$661 & 0.3$\pm$0.2\\
\CIV &  \nodata & 6978$\pm$1066 & 10.4$\pm$1.4\\
\CIV &  \nodata & 1431$\pm$480  & 3.4$\pm$1.2\\
\hline
\end{tabular}
\tablenotetext{a}{The redshifts of all emission lines are linked.}
\tablenotetext{b}{The line width of \lyb\ is fixed to be the same as that of the broad }
\end{table}
Assuming larger velocity dispersion values will decrease the limits and vice versa, because the observed spectrum places strong constraints on the Gaussian component.  For example, for $\sigma=  400$~\kms, the corresponding 68\%, 90\%, and 99\% confidence limits are 0.45, 0.75, and 1.6$\times10^{15}$~\cmsq.  For $\sigma=  100$~\kms, the corresponding 68\%, 90\%, and 99\% confidence limits are 1.0$\times10^{16}$, 1.0$\times10^{17}$, and  2.5$\times10^{19}$~\cmsq.
After searching for other absorption lines in the spectrum, we identified four candidates (Abs1--4) at wavelengths 1872.3\AA, 1917.7\AA, 2226.8\AA, and 2265.8\AA, respectively.  Possible origins of these lines include metal absorption lines associated with the lens galaxy, our own Galaxy, quasar host galaxy, or other \lya\ absorbers along the line of sight.\\
We first excluded Galactic absorption lines because the wavelengths do not match any strong UV ISM absorption lines \citep{blades88}.
We next considered absorption lines in the quasar host galaxy, in particular a series of \FeII\ absorption lines with rest-wavelengths shorter than that of \lya\ will fall in the observed wavelength range; however, they also do not match the wavelengths of the absorption line candidates.  
Thus, we concluded that Abs1--4 are due to some intervening systems, either the lens galaxy or other intervening clouds.  
Considering the redshift of the lens ($z_l \simeq 0.439$), Abs1 and Abs2 can be identified as the \OI$\lambda\lambda1302$ and \CII$\lambda\lambda1334$ absorption lines from the lens galaxy.
Abs3 is located at the wavelength of the expected \CIV$\lambda\lambda1549$ absorption line from the lens galaxy; however, since we do not expect to have high ionization ions in cold/warm ISM with $T < 5000 K$, we excluded \CIV\ from the lens as the interpretation of Abs3 and think it is possibly a \lya\ line from a different redshift at $z=0.8317$.
Abs4 does not correspond to any strong absorption lines in the lens, and we identified it as another \lya\ absorber at $z=0.8639$.
We align the \OI, and \CII\ absorption lines and the expected DLA from the lensing galaxy together in Figure~\ref{fig:zoom} to further illustrate the non-detection of the expected strong DLA feature. \\
We performed a final fit to the spectrum by freezing the continuum and emission line parameters and linking the wavelengths and FWHMs of Abs1 and Abs2 and, since they are both from the lens galaxy, and obtained a fitting statistic of $\chi^2/dof = 560.1/1082$.  We reported the best-fit parameters in Table~\ref{tab:ab}.  The spectrum, model, and residuals of this fit are shown in Figure~\ref{fig:model}.
We measured the redshift of the lens as $z_l = 0.4377\pm0.0007$, suggesting a multiplicative factor of 0.9971 or a linear shift of $1.71$\AA\ (observed frame) in the wavelength calibration difference between \hst/COS and Keck II LRIS compared to the \citet{myers99} redshift value.
\begin{table*}
\caption{Absorption Line Candidates in the COS UV Spectrum. \label{tab:ab}}
\begin{tabular}{ccccccccc}
\hline
\hline
Line & Wavelength & Identification & Redshift & Wavelength & FWHM & Velocity Dispersion & Optical & Column Density\\
& \AA & &  & \AA &\kms & (Rest) \kms & Depth & $10^{14}$~\cmsq\\
\hline
Abs1 & 1872.3$\pm$0.9 & \OI  & \nodata\tablenotemark{a} & \nodata\tablenotemark{a}    & \nodata\tablenotemark{a} & \nodata & 0.96$^{+0.98}_{-0.47}$ & 90\tablenotemark{b} \\
Abs2 & 1917.7$\pm$2.1 & \CII & 0.4377 & 1915.7$\pm$1.0 & 590$^{+450}_{-190}$ & $140_{-40}^{+260}$ & 0.41$^{+0.53}_{-0.35}$  & 30 \\
\nodata & \nodata     & \lya & \nodata\tablenotemark{a} & \nodata\tablenotemark{a}    & \nodata\tablenotemark{a} & \nodata & $<0.5$ & $<5.4$ \\
Abs3 & 2226.8$\pm$1.4 & \lya (or \CIV) & 0.8317 (or 0.4377) & 2226.8$\pm$3.3 & 280$^{+390}_{-270}$  & $<100$ & 0.98$^{+8}_{-0.48}$ & 8.2   \\
Abs4 & 2265.8$\pm$3.4 & \lya & 0.8639 & 2265.8$\pm$3.7 & 930$^{+1270}_{-540}$ & $200_{-60}^{+400}$ & 0.37$^{+0.45}_{-0.27}$ & 23 \\
\hline
\end{tabular}
\tablenotetext{a}{The redshifts, wavelengths, and widths of \OI\ and \lya\ lines at $z=0.4377$ are linked to those of the \CII\ line.}
\tablenotetext{b}{The spectral bin size of 1.5\AA\ is modeled in calculating the column densities.}
\end{table*}

To accurately access the significance of the detections of the \OI\ and \CII\ lines, we calculated a number of confidence intervals and report the one such that the corresponding optical depth is consistent with zero.
The \OI\ line is detected by $2.84\sigma$ (99.77\% one-sided probability) and the \CII\ line is detected by $1.21\sigma$ (88.69\% one-sided probability).  Considered jointly, the presence of NUV metal absorption lines from the lens galaxy is significant at the 99.975\% confidence (3.5$\sigma$) level. 
We also calculated the rest-frame velocity dispersions of the absorption lines by subtracting the spectral resolution of $\simeq$1\AA\ in quadrature, and report the values in Table~\ref{tab:ab}.  
Although having large uncertainties, the lens velocity dispersion is constrained to be $140^{+260}_{-40}\,\kms$, consistent with values of typical $L^*$ galaxies and the lens model value of \citet{asadi20}.

\section{Discussion}

We have analyzed a COS NUV spectrum of image A of the gravitational lens \bone.
We expected to find, based on the absorption seen in the X-ray spectrum, a DLA feature at the lens redshift with $\nh \sim 10^{20}\,\cmsq$ for image A \citep{dk09,chen13}.
We can place a 99\% upper limit on the neutral hydrogen column density with $\nh < 2\times10^{16}~\cmsq$.  We do detect weak \OI\ and \CII\ absorption lines associated with the lens galaxy at a combined significance of 3.5$\sigma$.
\bone\ has been well studied in the optical and X-ray bands, and there are multiple indicators showing that \bone\ contains a large amount of ISM, the optical extinction, X-ray absorption, and the presence of \MgII\ and \MgI\ absorption lines.  These previous measurements are consistent with our detection of \OI\ and \CII\ absorption lines.
The non-detection of the DLA feature is surprising, 
and thus, the lens galaxy of \bone\ is identified as a ``ghostly'' DLA, a DLA revealed by other absorption features but undetected \citep{jiang16, fath2016}.
Here, we discuss several possible explanations.

First, it is unlikely that the lens has an extremely high metal-to-gas ratio, such that we only detect absorption signatures from the metals or dusts including the extinction, X-ray absorption, and metal absorption lines.  Our 99\% \nh\ upper limit is four orders of magnitude below the expectation, and such an extreme metal-to-gas ratio is unprecedented.  Physically, collisions will always bind and mix the atom/molecules of difference species, and creating a region devoid of hydrogen atoms but with only metals is difficult. 
Second, it is also unlikely that most hydrogen atoms are in the ionized state because of the signatures of cold ISM detected, the \OI, \CII, \MgI, and \MgII\ lines.  In particular, the ionization potential for \MgI\ is 7.64\,eV lower than that for \HI\ (13.6\,eV) and the ionization potential for \OI\ (13.62\,eV) is comparable to that of \HI, while the O and Mg abundances are much lower.
The \OI\ column density is measured to be $\sim 9\times10^{15}\,\cmsq$ for image A, and for Solar metallicity, the expected \HI\ column density is $\sim 10^{19}\,\cmsq$, three orders of magnitude above our 99\% limit for a galaxy with $\sigma > 200$~\kms, and two orders of magnitude above the 90\% limit for a low mass galaxy with $\sigma=100$~\kms.  Thus, it is difficult to attribute the weak metal absorbers to a satellite galaxy associated with the lens.  
Image B has $\nh = 4.8\times10^{22}\,\cmsq$ from the X-ray spectrum, and this large amount of ISM is consistent with a late-type $L^*$ galaxy and is difficult to explain it in a satellite as well.

A more plausible explanation is that the expected DLA absorption trough is filled by the \lya\ emission from the lens galaxy. The lens emission contribution can be important because COS spectrum is slitless with a field-of-view of 1\sarc25 in radius, and emission from whole field of view will contribute to the spectrum, which includes the lens galaxy 1\sarc1 apart from image A.
\citet{fath2016} recently reported another ghostly DLA system at $z=1.70465$ along the line of sight of a $z=1.70441$ quasar, and the presence of DLA was also revealed by other metal absorbers.  In this case, the DLA and the quasar are very close, and the DLA trough is filled by the \lya\ emission from the quasar broad line region.  The authors also assumed that the ghostly DLA is falling towards the quasars, which yields an additional redshift such that its measured redshift is larger than that of the quasar.
In \bone, the lens galaxy is well separated from the background quasar, and the trough is presumably filled by the \lya\ emission from the lens galaxy, and thus the ghostly DLA in \bone\ represents a different population.
\lya\ emission from DLA systems has been detected \citep[e.g.,][]{joshi2016}, usually in the red wing of the DLA absorption profile.  The selection method for typical DLAs, however, will miss the \bone-like ghostly DLAs.  The \lya\ emission usually has a double-humped emission profile to escape from the galaxy \citep[e.g.,][]{ho2006,dijkstra2014}, and in \bone\ it is possible that the DLA trough is filled by the wings of the double-humped emission profile.  Depending on the detailed scattering conditions, the two peaks can be separated by up to a few or more Gaussian widths \citep[e.g., Figure\,5 of][]{dijkstra2014}, and the total profile can be broad and shallow.  The estimated \lya\ luminosity to fill the absorption trough is $4\times10^{42}\,\lumin$, which is close to the break luminosity of the \lya\ emitter luminosity function \citep[e.g.,][]{dressler15}. 
We note that the trough is not completely filled by the \lya\ emission, since there is signature of weak absorption at the line center.
A higher signal-to-noise ratio and resolution or slit spectrum excluding the majority of lens contribution of \bone\ is needed to investigate the details.

Regardless of the explanations, the expected DLA is not detected in \bone, where a large amount of \HI\ is expected.  DLAs are important probes to study the baryon distribution in the high redshift universe \citep[e.g.,][]{wolfe05}.  Therefore, it is important to characterize the statistical properties of the \bone-like ghostly DLAs to evaluate the potential biases introduce by these systems.

\section*{Acknowledgements}
We acknowledge C.\,S.\ Kochanek for helpful comments, C.\ Fassnacht for providing more details on the Keck II LRIS spectrum of \bone, and the anonymous referee for additional beneficial suggestions.
Support for the program, HST-GO-13283, was provided by NASA through a grant from the Space Telescope Science Institute, which is operated by the Association of Universities for Research in Astronomy, Inc., under NASA contract NAS 5-26555.

\bibliographystyle{mnras}

\begin{thebibliography}{}
\bibitem[Asadi et al.(2020)]{asadi20} Asadi, S., Zackrisson, E., Varenius, E., et al.\ 2020, \mnras, 492, 742

\bibitem[Bohlin, Savage \& Drake (1978)]{bsd78} Bohlin, R. C., Savage, B. D., \& Drake, J. F. 1978, \apj, 224, 132

\bibitem[Blades et al.(1988)]{blades88} Blades, J.~C., Wheatley, J.~M., Panagia, N., et al.\ 1988, \apj, 334, 308 

\bibitem[Chen et al.(2013)]{chen13} Chen, B., Dai, X., Kochanek, C.~S., \& Chartas, G.\ 2013, arXiv:1306.0008 

\bibitem[Dai et al.(2003)]{dai03} Dai, X., Chartas, G., Agol, E., Bautz, M.~W., \& Garmire, G.~P.\ 2003, \apj, 589, 100 

\bibitem[Dai \& Kochanek(2005)]{dk05} Dai, X., \& Kochanek, C.~S.\ 2005, \apj, 625, 633 

\bibitem[Dai et al.(2006)]{dai06} Dai, X., Kochanek, C.~S., Chartas, G., \& Mathur, S.\ 2006, \apj, 637, 53 

\bibitem[Dai \& Kochanek(2009)]{dk09} Dai, X., \& Kochanek, C.~S.\ 2009, \apj, 692, 677 

\bibitem[Dijkstra(2014)]{dijkstra2014} Dijkstra, M.\ 2014, \pasa, 31, e040 

\bibitem[Dressler et al.(2015)]{dressler15} Dressler, A., Henry, A., Martin, C.~L., et al.\ 2015, \apj, 806, 19 

\bibitem[El{\'{\i}}asd{\'o}ttir et al.(2006)]{ela06} El{\'{\i}}asd{\'o}ttir, {\'A}., Hjorth, J., Toft, S., Burud, I., \& Paraficz, D.\ 2006, \apjs, 166, 443 

\bibitem[Falco et al.(1999)]{falco99} Falco, E.~E., Impey, C.~D., Kochanek, C.~S., et al.\ 1999, \apj, 523, 617 

\bibitem[Fathivavsari et al.(2017)]{fath2016} Fathivavsari H., Petitjean P., Zou S., Noterdaeme P., Ledoux C., Krühler T., Srianand R., 2017, \mnras, 466, L58

\bibitem[Green et al.(2012)]{green12} Green, J.~C., Froning, C.~S., Osterman, S., et al.\ 2012, \apj, 744, 60 

\bibitem[Hansen \& Oh(2006)]{ho2006} Hansen, M., \& Oh, S.~P.\ 2006, \mnras, 367, 979 

\bibitem[Jiang et al.(2016)]{jiang16} Jiang, P., Zhou, H., Pan, X., et al. \ 2016, \apj, 821, 1

\bibitem[Joshi et al.(2017)]{joshi2016} Joshi, R., Srianand, R., Noterdaeme, P., \& Petitjean, P.\ 2017, \mnras, 465, 701 

\bibitem[Mediavilla et al.(2005)]{med05} Mediavilla, E., Mu{\~n}oz, J.~A., Kochanek, C.~S., et al.\ 2005, \apj, 619, 749 

\bibitem[M{\'e}nard \& Chelouche(2009)]{menard09} M{\'e}nard, B., \& Chelouche, D.\ 2009, \mnras, 393, 808 

\bibitem[Motta et al.(2002)]{motta02} Motta, V., Mediavilla, E., Mu{\~n}oz, J.~A., et al.\ 2002, \apj, 574, 719 

\bibitem[Myers et al.(1999)]{myers99} Myers, S.~T., et al.\ 1999, \aj, 117, 2565 

\bibitem[Mu{\~n}oz et al.(2004)]{munoz04} Mu{\~n}oz, J.~A., Falco, E.~E., Kochanek, C.~S., McLeod, B.~A., \& Mediavilla, E.\ 2004, \apj, 605, 614 

\bibitem[Nadeau et al.(1991)]{nadeau91} Nadeau, D., Yee, H.~K.~C., Forrest, W.~J., et al.\ 1991, \apj, 376, 430 

\bibitem[Oguri \& Marshall(2010)]{om10} Oguri, M., \& Marshall, P.~J.\ 2010, \mnras, 405, 2579 

\bibitem[Prochaska \& Herbert-Fort(2004)]{ph04} Prochaska, J.~X., \& Herbert-Fort, S.\ 2004, \pasp, 116, 622 

\bibitem[Rusin et al.(2002)]{rusin02} Rusin, D., Norbury, M., Biggs, A.~D., et al.\ 2002, \mnras, 330, 205 

\bibitem[Toft, Hjorth \& Burud (2000)]{toft00} Toft, S., Hjorth, J., \& Burud, I. 2000, \aap, 357, 115 

\bibitem[Wolfe et al.(2005)]{wolfe05} Wolfe, A.~M., Gawiser, E., \& Prochaska, J.~X.\ 2005, \araa, 43, 861 

\bibitem[Wucknitz et al.(2003)]{wucknitz03} Wucknitz, O., Wisotzki, L., Lopez, S., \& Gregg, M.~D.\ 2003, \aap, 405, 445
\end{thebibliography}

\bsp	
\label{lastpage}
\end{document}